# STATIC INTERNAL REPRESENTATION OF DYNAMIC SITUATIONS REVEALS TIME COMPACTION IN HUMAN COGNITION

José Antonio Villacorta-Atienza[1,2], Carlos Calvo-Tapia[2,*], Sergio Díez-Hermano[1,*], Abel Sánchez-Jiménez[1,2], Sergey Lobov[3], Nadia Krilova[3], Antonio Murciano[1], Gabriela López-Tolsa[4], Ricardo Pellón[4], Valeri Makarov[2,3]

1. Biomathematics Unit (BEE Department), Faculty of Biology, Complutense University of Madrid, Spain.
2. Institute of Interdisciplinary Mathematics (IMI), Faculty of Mathematics, Complutense University of Madrid, Spain.
3. Lobachevsky State University of Nizhny Novgorod, Russia.
4. Department of Basic Psychology, Universidad de Educación a Distancia (UNED), Spain

**The time-changing nature of our world demands processing of huge amounts of information in fast and reliable way to generate successful behaviors [Llinás 2001]. Therefore, significant brain resources are devoted to process spatiotemporal information [Rao 2001, Livesey 2007, Kraus 2013]. Neural basis of spatial processing and their cognitive correlates are well established mostly for static environments [O'Keefe 1976, Fynn 2004, Epstein 2017]. Nonetheless, in time-changing situations the brain exploits specific processing mechanisms for temporal information based on prediction and anticipation [Bubic 2010], as time compression during visual perception [Ekman 2017] and mental navigation [Arnold 2016]. Alternative hypothesis of time compaction integrates both views, postulating that dynamic situations are internally represented as static spatial maps where temporal information is extracted by predicting and structuring the relevant interactions [Villacorta-Atienza 2010]. Nevertheless, empirical approaches tackling the biological soundness of time compaction are still lacking. Here we show that performance in a discrimination learning task involving dynamic situations can be either favored or hampered via previous exposition to interfering static scenes. In this sense, men were effectively conditioned in contrast to a control group, in coherence with the hypothesis. Meanwhile, women performed on par with control men, regardless of the previous conditioning. This suggests time compaction is a salient cognitive strategy in men when dealing with dynamic situations, while women seem to rely on a broader range of information processing strategies [Peña 2008, Picucci 2011]. Finally, we further corroborated the time compaction mechanism involved in these experimental findings through a mathematical model of the experimental process. Our results point to some form of static internal representation mechanism at cognitive level involved in decision-making and strategy planning in dynamic situations. The existence of time compaction in the human brain could provide a functional framework unifying essential aspects of cognition demanded for active interaction with our world, introducing a new venue to embed human-like basic cognition in robots [Villacorta-Atienza 2013].**

Temporal and spatial dimensions are prominent organizing features in nature, so anticipation to complex dynamic hazards is mandatory to survive [Maldonato 2012]. Thus, prediction has been proposed as the ultimate brain function [Llinás 2001]. Founded on this capability, time compaction provides an efficient mechanism to structure and process the critical information contained in dynamic environments. According to this theory, a dynamic situation is internally represented by the predicted interactions among its elements (including, if pertinent, the subject itself). Such interactions, considered as the relevant events, are spatially structured in a static map. Therefore, compacting time results in a static spatial internal representation of the perceived time-changing situation [Villacorta-Atienza 2010]. As time is no present, the amount of information the brain processes is drastically reduced. This way, the presence of time compaction in the human brain suggests that the cognitive mechanisms involved in both time-changing experiences and static scenarios could be closely related, opening new opportunities to the study of these cognitive phenomena [Villacorta-Atienza 2013].

We assessed time compaction carrying out a decision-making experiment, where human participants were prompted to classify different situations displayed in a computer screen, according to an underlying rule. In order to introduce the experimental procedure, let us consider: i) a simple dynamic situation where two circles are moving with collision trajectories, and ii) a static scene with a single immobile circle at the same location as the collision area in the dynamic situation. According to the hypothesis of time compaction, the dynamic situation would be internally represented by the collision area (relevant event). Thus, the static scene, denoted by SM (Static Matching scene), and the dynamic situation, or DM (Dynamic Matching situation), would have similar static internal representations (Fig. 1A). During the experiment, properly classifying SM could either improve or worsen later classification of DM. This connection was exploited to unveil the presence of time compaction through a discrimination learning task. Participants go through two sequential phases: a conditioning training phase and a testing game. In the first phase a succession of consecutive static scenes is displayed on the screen. Each scene randomly shows one out of three different configurations: red circle in the center (SM) or on the left/right (non-SM) in relation to a reference green circle. The subject should press either up or down arrow key on each presentation of a scene, and receives a feedback: "success" if the pressed key matches a pre-established underlying rule (linking scene and key) or "fail" otherwise. At the beginning the subject does not know the correct association of keys and situations. Then, by trial-and-error he/she learns it and eventually begins responding correctly. Once the rule has been learnt, the testing game starts. During the game the same scheme is followed, in this case showing dynamic situations. The subject should figure out the rule linking the up and down keys with six different dynamic situations, two DMs (collision; Fig 1B, highlighted) and four non-DM (no collision; Fig 1B, non-highlighted), displayed in random sequence. Note that the subject watches only the initial part of a situation, i.e., collision is not shown on the screen.

Under the time compaction hypothesis, the previous learning of the SM-key association during the training should interfere the rule learning for DMs during the game. Through this relation we established favorable or adverse conditionings in terms of time compaction by adapting the underlying rule. We considered three experimental groups: two including participants whose learning performance would be either favored or hampered, and a third control group that would serve as a baseline of non-influence (Fig. 1C). In the Favored group, the sequence comprised one SM and two non-SMs. The SM arrow key coincides with the one associated to DM during the game, while non-SM and non-DM were both linked to the opposite arrow key. This was supposed to favor learning of the association rule, decreasing the number of trials required to finish the game. The same scheme was preserved for the Control group but SM was replaced by a third non-SM, removing the postulated effect of time compaction. This way, conditioning and game phases were independent. Finally, in the Hampered group the SM arrow key was opposite to the one associated to DM, and non-SM keys were mismatched to avoid spurious associations. This was expected to hamper the learning of the game's logic. It should be remarked that prior to the experiment, the participants were not aware either of the existence of collision-related dynamic situations and of the relationships between static scenes and dynamic situations displayed in both stages (see Methods). This way, the experiment was intended to uncover time compaction as an unconscious mechanism to process dynamic situations, revealing 1) interactions as salient, and so relevant, events, and 2) that such events are represented as static structures.

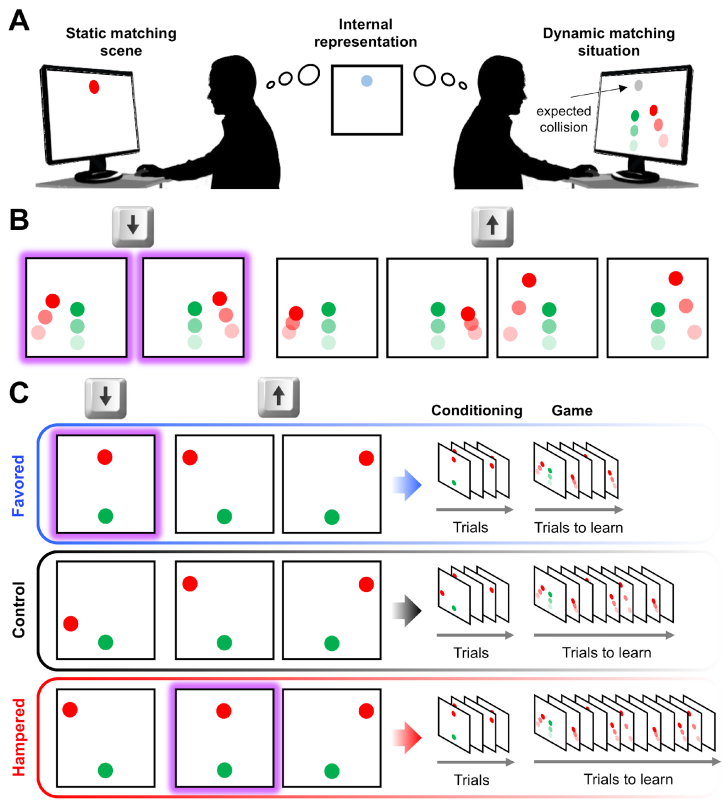

**Figure 1. Experimental approach for testing time compaction hypothesis. A.** According to time compaction, static matching scene, or SM (left screen), and dynamic matching situation, or DM (right screen), have similar internal representations since position of immobile circle in SM matches with location of expected collision in DM. **B.** The testing game consists in figuring out, by trial-and-error, the relationship between up-down arrow keys and six dynamic situations: the DM, two non-DMs, and their symmetric counterparts. In DM situations (highlighted in purple and linked to the down arrow key) the red and green circles move to collide, though collision is not shown to the participant. Non-DM situations (linked to the up arrow key) show circles whose velocities prevent collision. Green circle's velocity remains constant in all situations. Faint colors represent past frames in the movement of the circles. **C.** Different relations established in the conditioning training phase between static scenes and up-down arrow keys could favor or hamper the testing game performance. For Favored/Hampered groups the SM scene (highlighted in purple) is associated to the same/different key than in DM. For Control group, no SM is displayed, so the rule is not connected to DM according to time compaction hypothesis. Assuming Control group is not conditioned, the Favored/Hampered groups should need less/more trials to learn the game rule. Note that the relationship between arrow keys and displayed scenes and situations is randomly counterbalanced for each participant.

Considering the set of answers from all participants (126 men and 135 women) we characterized the population learning process by defining the *success rate* per trial (Fig. 2A). The analysis of the population learning process was performed by modeling the success rate through general estimating equations (GEE) and logistic regression analysis. The independent variables involved, apart from trial (inherent to the learning), were gender, group and researcher (the person who conducted the experiment). Only gender and group were found significant. Population learning in men was significantly faster/slower for Favored/Hampered groups compared with Control. Nonetheless women showed no significant differences either among Favored, Hampered and Control groups or against men in Control group.

To characterize the individual learning performance, we introduced the *learning length* of each participant as the number of trials required to figure out the rule (Fig 2B). Men from Favored/Hampered groups had significantly higher/lower probability of learning at early trials than men in Control group. On the contrary, conditioning phase did not affect the individual learning of women, showing no significant differences against men in Control. These differences in individual learning performance are evidenced through Kaplan-Meier curves and survival analysis (Fig 2B, inset panels). These findings reveal an intrinsic relationship between SM scenes and DM situations. That suggests men primarily represent dynamic situations by the static representation of relevant interactions (DMs and its SM correlate, which contains the location of the expected collision), thus significantly reducing the information complexity. Furthermore, our results support women seemingly rely on a broader range of cognitive strategies [REF Peña 2008, Picucci 2011] in conjunction with time compaction, as revealed by non differences with Control men.

In order to go deeper into participants' decision making, we aimed to analyze potential relationships between verbalization of the game association rule and the individual learning performance. For this purpose, after they finished the game, participants were asked for writing the rule they figured out. Generalized linear model (GLM) and logistic regression revealed that fast-learning men showed a significant tendency to answer in terms of 'collision', regardless the experimental group; otherwise, the more learning time they required, the more they seemed to resort to an alternative strategy based on other criteria. On the contrary, for women such a tendency was not observed (Fig. 2C, men and women upper panels). This is coherent with previous results, pointing to gender differences in the preference for time compaction-based strategy for learning the game rule.

Additionally, we studied other factors that might influence the game performance. To this extent, we analyzed the answer latency, i.e. the time elapsed since the displayed situation disappeared until the participant pressed the key. GEE analysis revealed that answer latency did not significantly differ either between men and women or among experimental groups (Fig. 2C, men and women bottom panels). This indicates all participants take comparable times to resolve individual trials, which suggests similar difficulty for figuring out the rule regardless previous conditioning. Thus, as postulated by time compaction, differences in game performance could be attributed to the effect of the conditioning on the information features used to solve the game.

The experimental results here reported suggest a time compaction-based mechanism involved in human decision-making when dealing with dynamic situations. To corroborate such a mechanism from a complementary perspective, we mathematically modeled the learning during the game under conditioning and control conditions, and compared the theoretical and experimental results. The model assumes a successful answer probability with recalling exponential decay [Nembhard 2001, Averell 2011], biased according to time compaction premises: due to the assumed similar internal representation of SM and DMs, favorably conditioned participants started the game with a set of associations already learned, whereas hampered participants needed to re-elaborate the game's rule (Supplementary Material).

The modeling of the learning process during the game involved the population learning process, described by analytical equations, and the individual learning performance, via computational simulation, engaged in a recursive process. The analytical model describes the probability of successful answer at each trial $T$, which is given for Favored, Control and Hampered groups respectively, by:

$$P_F = 1 - \frac{1}{3}\left(\frac{5}{6}\right)^{T-1}[1 + \alpha(a,T) + \beta(b,T)],$$

$$P_C = 1 - \frac{1}{2}\left(\frac{5}{6}\right)^{T-1}[1 + \alpha(a,T) + \beta(b,T)],$$

$$P_H = 1 - \frac{1}{2}\left(\frac{5}{6}\right)^{T-1}\left[\frac{4}{3} + \alpha(a,T) + \beta(b,T)\right],$$

where α and β stands for the recalling terms for the second and third appearance of a stimulus, and *a* and *b* denote the corresponding recalling constants (see Supplementary).

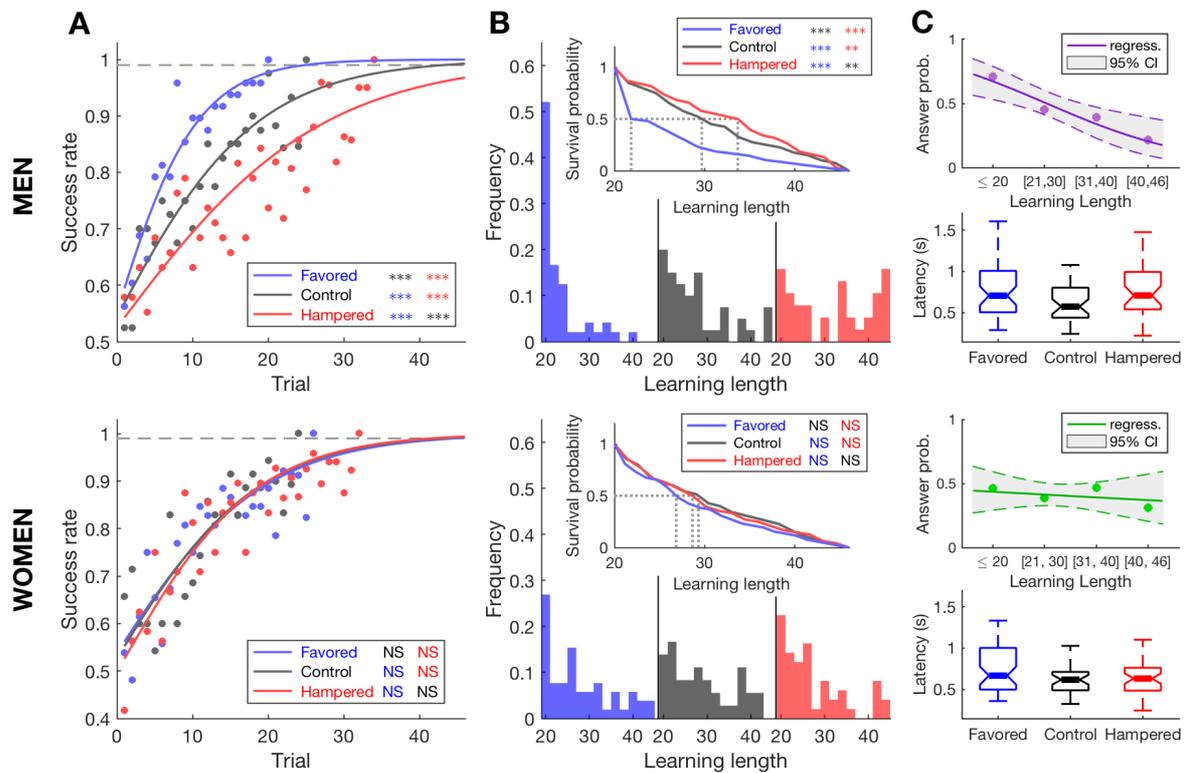

**Figure 2. Population learning process and individual learning performance.** Top and bottom rows correspond to men and women, respectively; blue, black, and red colors stand for Favored (F), Control (C), and Hampered (H) groups respectively (F: 48 men, 52 women; C: 40 men, 35 women; H: 38 men, 48 women). **A.** Population learning process as success rate per trial (sample average of correct answers) was considered up to 0.99 (dashed lines). Men in Favored/Hampered groups learned the game association rule faster/slower than men in Control group (F vs. C: $p$ = 3e-04; F vs. H: $p$ = 9.8e-13; C vs. H: $p$ = 2.5e-04). Population learning process in women showed no dependence on the direction of conditioning (F vs. C: $p$ = 0.43; F vs. H: $p$ = 0.28; C vs. H: $p$ = 0.93). Men from Control group and women from all groups did not show significant differences ($p$ = 0.23, $n$ = 40 vs. 135). Curves describe the logistic regression of the corresponding population learning. **B.** Individual learning performance. The learning length distribution shows it is significantly more/less likely for men in Favored/Hampered groups to learn the association rule at early trials than in Control condition (F vs. C: $p$ = 7.2e-04; F vs. H: $p$ = 3.5e-8; C vs. H: $p$ = 5.8e-03). This effect was not observed in women (F vs. C: $p$ = 0.98; F vs. H: $p$ = 0.67; C vs. H: $p$ = 0.76). Men from Control group and women from all groups did not show significant differences ($p$ = 0.95, $n$ = 40 vs. 135). Insets show the Kaplan-Meier curves used to quantify differences in individual learning performance [Smith 2011]. Dotted curves denote the learning length where learning probability is 0.5. **C.** Rule verbalization and answer latency. Men and women top panels: rule verbalization in terms of 'collision' had a probability above 0.7 for those men who quickly learned the rule, regardless the direction of the conditioning, and decreases with the learning length. No dependency was observed in women. Curves and grey areas denote logistic regression and mean confidence intervals at 95% respectively (men: $p$ = 1.2e-4, $n$ = 126; women: $p$ = 0.67, $n$ = 135). Learning lengths were grouped in four groups: those with learning lengths lower than or equal to 20 and those with learning lengths in [21,30], [31, 40], and [40, 46]. Men and women bottom panels: the answer latency showed no differences among groups (GEE regression, F vs. C: $p$ = 0.13, $n$ = 100 vs. 75; F vs. H: $p$ = 0.45, $n$ = 100 vs. 86; C vs. H: $p$ = 0.42, $n$ = 86 vs. 75) and for gender ($p$ = 0.4, $n$ = 126 men vs. 135 women). ***: < 0.001; **: < 0.01; *: < 0.05; NS: No Significant difference.

The recursive process to simulate the learning during the game phase is as follows. First, the analytical model of the population learning process for Favored, Control and Hampered groups was fitted to the corresponding experimental data. The two parameters obtained from the fitting were then incorporated into the computer simulation to get those participants who could not follow the model, which were extracted from the sample (though they may eventually be included again in subsequent iterations). The model is then fitted to the population learning data from the updated sample and the cycle begins again. This recursive process continues until the sequence of parameter values converges. The final parameter values were introduced into the computer simulation to create a set of virtual participants solving the game. The learning length distribution of this virtual sample is compared with the experimental one for the final updated sample in Fig. 3. The agreement of the experimental data with the model provides additional support to the proposed mechanism based on time compaction, particularly considering that for Favored and Control groups only 4% and 5% of the sample respectively were unfit to follow the model. Modeling of Hampered group showed a 24% of subjects who did not behave according the model. This subpopulation corresponded to participants with high learning lengths, indicating that the model does not capture properly specific behaviors of those subjects, possibly related with model assumptions (e.g. hampered participants could require more than three stimulus appearances to learn, i.e. more than two parameters in the model) or exogenous factors (motivation, frustration, etc. not included in the model; see Supplementary).

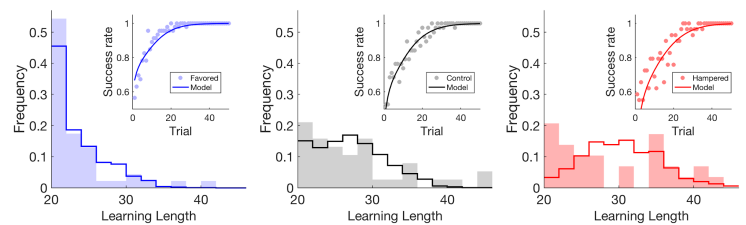

**Figure 3. Mathematical model of game process based on time compaction.** **A.** Individual learning performance was computationally modeled with the parameters obtained after convergence of the recursive process for modeling the learning during the game (see main text). Main panels: histogram and stair-like curve illustrate experimental and modeled individual learning performance respectively (Favored in blue, Control in black, and Hampered in red). Insets: corresponding theoretical population learning process (curves) fitted to the experimental data of the updated sample after convergence (dots).

The presence of time compaction revealed by the proposed discrimination learning game has uncovered a clear gender biasing. Though time compaction is hypothesized as a sex independent cognitive mechanism, gender differences here reported are in full agreement with the potential role of time compaction as cognitive strategy. On the one hand, men frequently use holistic strategies, gathering information about spatial relations among stimuli to plan in advance. In contrast, women more often use segmentary strategies, concentrating on particular aspects of relevant elements and less focused on their relationships [Peña 2008]. This supports men could exploit time compaction as a prominent strategy for dealing with dynamic scenarios in a fast and reliable way, a critical feature theoretically postulated for this mechanism [Villacorta-Atienza 2010]. These gender differences match with gender-linked pressure for traits as foraging division (hunting versus gathering) or agonistic behaviors (projectile throwing and intercepting) [Eals 1994, Jones 2003]. This selection process [Ecuyer-Dab 2004] suggests a sex-related enhancement of specific biological mechanisms involved in complex dynamic interactions, like hunting or fighting, consistent with the nature of time compaction and its potential impact over the processing of time-changing experiences in the human brain. Actually, gender differences described here, together with the associative nature of the experiment conducted, point time-compacted static representation of dynamic situations relays at cognitive level, beyond visual substrate [Schutz 2007, Kerzel 2003], where mechanisms as time compression [Ekman 2017] and visual extrapolation [Delle Monache 2017, Russo 2017] are located. In this regard, the notion of time compaction resonates with the concept of cognitive map [Epstein 2017]. Cognitive map theory predicts hippocampus is involved into navigation to unmarked locations, defined by their relationships with the environment, but not into navigation to visible static goals [Morris 1982]. Nevertheless avoidance of visible dynamic objects depends on the hippocampus [Telensky 2011]. This apparent contradiction suggests dynamic situations are internally represented from the relationships among perceived elements. This is coherent with time compaction, where the static representation is generated from the expected interactions among elements in the time-changing environment, including, if relevant, the subject itself. Moreover, it raises the possibility that cognitive maps and time-compacted static representations share neurobiological mechanisms. This could open new perspectives for understanding the neural substrate enabling cognitive interaction with our dynamic world, and introduce new venues to embed human-like basic cognition in robots [Villacorta-Atienza 2013].

## Methods

### Experimental procedure

Participants were subjected to a computer-based trial and error training consisting of two phases, one static conditioning followed by one dynamic game (see **Fig. 1**). They completed the procedure individually, as no interaction between them was allowed. Before the session the researchers read aloud the instructions to be followed, which were also shown in the screen. Any doubt asked by the participants was answered citing again the instructions in the monitor. The instructions referred to conditioning and game phases as static and dynamic stages, so no mention to the relationship between both phases was done to the participants. The structure of the instructions was as follows. First, it was detailed that the task was an associative experiment consisting on a static phase and a latter dynamic phase, whose objective was to find out the relationship between the displayed images and the up-down arrow keys; displayed images were shown for 1.5 s and were two static or dynamic circles (diameter 1 cm), colored in red and green, and presented on a white background. Simple circles were chosen to avoid prior cognitive bias [Reed 1996]. The instructions relating to the static phase (conditioning) specified the green circle was located at the bottom center of the image whereas the red circle appeared randomly, from trial to trial, at the upper left, center or right of the screen. On the other hand, the instructions concerning the dynamic stage explained that both green and red circles followed straight lines at constant velocity; while the green circle always started from the bottom center and moved upwards vertically, the red circle changed its movement randomly from trial to trial, and could appear at the bottom left or the bottom right of the screen. Therefore, during instructions reading the participants were not aware about possible collisions between the moving circles.

During the game the red circle moved diagonally alternating between three different velocities; green circle's velocity: 4.5 cm/s, red circle's velocity: 1/3, 2/3, and 4/3 relative to green circle's speed. Dynamic situations where velocity ratio was equal to 2/3 correspond to circle collision, i.e. to the DMs. In DMs, collision would take place 1 second after circles disappeared (the image was displayed during 1.5 seconds).

There was no time limit to complete the stages, even though there was a maximum number of 80 possible trials per phase (conditioning and game). The learning criterion was 18 correct responses in the last 20 trials for each participant. The trial in which participants reached the criterion was considered the learning length. Note that 20 trials is the most frequent number of trials needed for every scene to appear at least twice **(Fig. S1**; see Supplementary Material for further insight into task complexity and stimulus difficulty).

After both conditioning and game phases, participants were prompted to write down the underlying association rule they thought best explained the link between the displayed situations and the arrow keys. At the end of game phase they were asked to fill a form in order to gather information about their age and gender.

The task was programmed in MATLAB (MathWorks) v17.

### Sample description

Experiments were conducted over randomly composed groups of men and women. The final sample comprised those participants who fulfilled the task in both phases showing a stable performance (see Supplementary Material and Extended Data), which included those participants with learning lengths smaller or equal to 46 (n=261, 82% of the total sample; w: 135, m: 126).

Mean age of volunteers was 21.63 (SD: 7.23) for women and 23.08 (SD: 8.86) for men (no difference, Welch test p-value = 0.11). None reported prior attentional problems. All had normal or corrected-to-normal vision. All were naïve to the study's purpose and had no experience with the tasks and stimuli used here. Most subjects finished the experiment within 10 minutes.

### Statistical analysis

Analysis of experimental results focused on the dynamic phase.

*Time-to-event curves estimation*

A time-to-event function representing the probability that a certain proportion of individuals had learned at a given trial was obtained through the Kaplan-Meier non-parametric estimator. To assess whether gender and experiment might be simultaneously affecting the time-to-event curves estimation a multivariate Cox proportional hazard regression model was fitted (Jan-Eimermacher 2011, Smith 2011). The model assumptions were checked via the Schoenfeld test (Schoenfeld 1983, Abeysekera 2009) and residuals plots against time for each covariate (**Fig. S2A-B**). This helped to identify two potential cutoffs at learning lengths 20 and 46 that might be hampering the proportional hazards assumption. We extended the Cox model to allow for time-dependent coefficients (Saegusa 2014) using a step function, dividing the sample in three learning length intervals: less than 20, 20-46, and greater than 46 (**Fig. S2C-D**). The revised fit revealed that the effect of the experiment was essentially limited to the first two intervals.

*Learning curves estimation*

General estimating equations (GEE) were used to model the probability of success at a given trial, as it is risky to consider trials independent from each other and GEEs allow for different correlation structures (Hanley 2002, Hardin 2005, Hin LY 2009). Binomial "logit" was used as link function given the dichotomous nature of the response variable (either fail a trial or not). A backwards stepwise elimination procedure was followed to select the minimal set of variables with significant explanatory power. Variables taken into account were gender, group and trial number. Successive nested models were compared using the F-test and the Quasilikelihood Information Criterion (QIC) (**Table S1**) (Pan 2001). Interaction terms were interpreted separately for gender and experiment (**Table S2**). To ensure the model was characterizing the learning phase, such a stage was considered as the set of trials with mean success rate lower or equal to 0.99. A factor indicating the different researchers that conducted the experiment was introduced in the model and no significant effect was found (F-test p-value = 0.21). This is indicative that there was no detectable bias introduced in the results by the researcher and confirms the replicability of our experimental approach.

*Association rule verbalization and latency times*

Potential relationships between the verbalization of the association rule found by the participants and the learning length were explored by using a Generalized Linear Model (GLM) with binomial "logit" link function. Regarding verbalization, 96.2% of participants (251 from 261) expressed the dynamic rules in two main categories: in 'collision' terms (containing words as collision, crash, finding, etc.) or in 'velocity' terms (with words as velocity, speed, etc.). The rest of participants (3.8% – 10 from 261) wrote the dynamic rule by simultaneously including 'collision' and 'velocity' terms or by means of spatially related descriptions based on directions, positions, etc. On the other hand, the learning lengths were discretized into four groups: those participants who learnt before or at trial 20, and those with learning lengths in [21,30], [31, 40], and [40, 46]; this sample distribution allows the proper conditions for the statistical analysis. Latency time differences through trials due to experiment or gender were checked by fitting a GEE with gaussian "identity" link function, with latency as a continuous response variable.

All statistical analyses were performed in R v3.3.1, using the packages survival (Therneau 2015), survminer (Kassambara 2017), geepack (Hõjsgaard 2006), stats, base (R Core Team 2016), and dplyr (Wickham 2017).

### Mathematical model

The mathematical model describes the experimental process of figuring out the underlying association rule during the testing game. The model quantifies the probability of successful answer at each trial, based on four assumptions: 1) the first time a stimulus appears its association key will be learnt (if the pressed key was not correct the opposite one will be the right association key); thus wrong answers when the same stimulus appears again will be due to defective recalling, 2) the probability of recalling the key associated to a specific stimulus decays exponentially with time, 3) the recalling decay rate will depend on the number of times the same stimulus has appeared and will not depend on the specific stimulus, and 4) we assumed that after the same stimulus has appeared four or more times the association key recall probability will be 1 and the selected key will be always correct. This is a reasonable assumption since the successful answer probability once the same stimulus has appeared four or more times is 0.97 and 0.88 for men in Favored and Hampered groups respectively, and 0.91 considering together women and Control men.

Time compaction was introduced into the model in terms of the associations (between displayed situations and keys) to be learned during the game after the previous conditioning training. For modeling the performance of the Favored group during the game, only the four associations for non DMs had to be learned, since it is assumed the remaining two associations, for DMs, were previously learned in the conditioning training phase. On the contrary, modeling the Hampered group performance assumes the two associations for DMs must be re-elaborated during the game, since they were wrongly learned during the conditioning training phase.

The model comprises the analytical description of the population learning process and the computational simulation of the individual learning performance, and depends on two parameters, the recalling decay rates, (see Supplementary). The learning of the associative rule during the game was modeled by a recursive process where, in each iteration, the experimental sample is updated to reject those participants (from each group) that could not follow the model. First, the model parameters for each experimental group were obtained by fitting the model equations to the experimental population learning data for the group sample. With these parameter values, it was checked which subjects could not follow the mathematical model. These participants were then extracted from the whole sample and the model was fitted again to the updated sample to obtain new parameter values. This way, the cycle was repeated recursively until the sequence of parameter values converged. Note that, in every recursive step, the removal of those subjects that could not follow the model, was done over the initial –whole- sample, so the same participant could be eventually eliminated from the sample in an specific iteration and included again in next iterations. After convergence, the final parameter values were introduced into the computational model to generate a set of $10^5$ virtual participants, simulating the game process, and obtaining the theoretical individual learning performance (stair-like curves in Fig. 3).

Probabilistic model parameters were estimated from the sample learning rates by trust-region-reflective least squares fitting. Previously each participant's answer vector was filled with ones (success) from its learning length trial to the maximum learning length (46 trials). Initially the model was fitted only to women and control men together. Bootstrapped confidence intervals were obtained for the two parameters. Then, the model was fitted to men in Favored and Hampered groups constraining the first recalling rate values to the interval range found in Control, as we expect that first repetition of any stimulus contain similar information among conditionings. Final values of the pair of recalling decay rates for Favored, Control, and Hampered groups were: $[a, b]$ = [0.2245, 0], [0.2244, 0.0251], and [0.2244, 0.1034] respectively.

The process to determine the subjects who did not follow the model consist of several steps. 1) For each subject we considered its sequence of stimulus, i.e. the sequence of dynamic situations displayed during its participation. 2) Since the length of this sequence was equal to the participant's learning length, to introduce these stimuli into the computational simulation of the game we completed the sequence, randomly adding stimuli until the maximum learning length was reached. This step was repeated $10^3$ times with each single stimulus sequence. 3) Each one of the previous completed stimulus sequence was then introduced 1000 times into the computational simulation to obtain their learning lengths. Thus, from the set of $10^6$ learning lengths, it was obtained the learning length frequency distribution for the specific subject assuming he follows the mathematical model during the game. Therefore, if the actual participant's learning length is out of the frequency distribution limits, the subject is extracted from the sample.

Model fitting, bootstrap and experiment simulation were programmed in MATLAB (MathWorks) v17 and R v3.3.1, package nlstools (Baty 2015).

# SUPPLEMENTARY MATERIAL

**Overview of data analysis**

The main objective of the proposed experiment is to figure out, by trial-and-error, the association rule underlying the testing game phase after the participant has solved the previous conditioning training stage. The answer of each participant was stored as a binary vector, 1/0 for right/wrong answer. Thus, the set of answers for each experimental group was arranged as a 1-0 matrix, where rows and columns correspond to participants and trials respectively. The specific analysis of each dimension allows quantifying the population learning process (collapsing information from rows) and individual learning performance (grouping information from columns)(Fig. 1S).

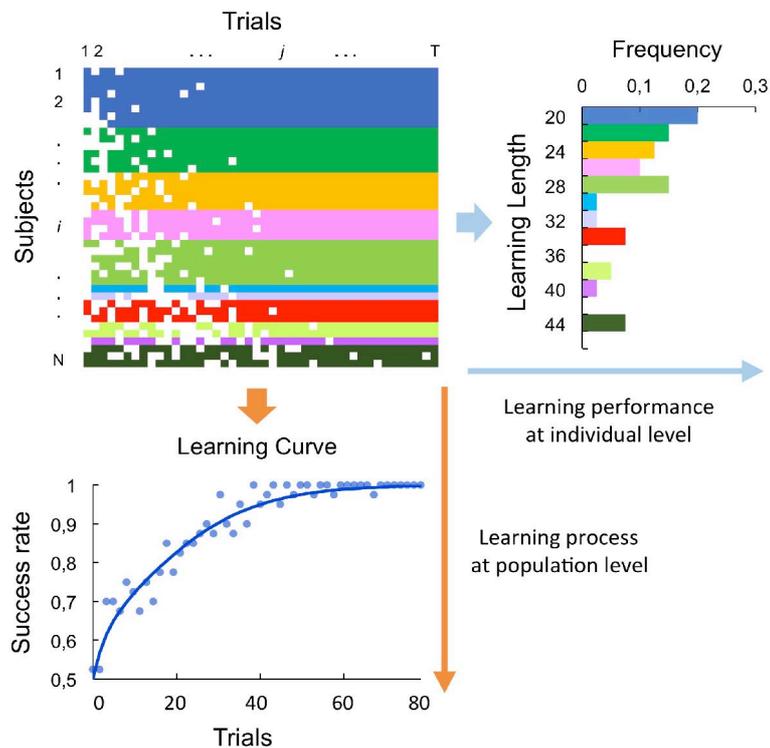

**Figure 1S. Data structure and analysis of data.** The experimental results are structured in a binary matrix of fails (white squares) and successes (colored squares), per subject and trial. In the above representation participants with the same learning length are grouped by color. Collapsing the rows dimension give a mean success rate per trial, enabling the analysis of the learning process at population level. Collapsing the columns dimension using predefined criteria (learning defined as 18 successes out of 20 trials) allows analyzing learning performance at individual level via the learning lengths frequencies.

**Task complexity**

The main task in the proposed experiment is to find, by trial-and-error process with feedback, the association rule between displayed static scenes/dynamic situations and up-down arrow keys. This task has been designed to exhibit low complexity, allowing any person could fulfill it with no difficulty. The first time an image is displayed the participant will press one of the two keys randomly (no previous information is available). However the feedback provided to his answer will lead to the learning of the association rule for the shown image: if the feedback confirms the correctness of the answer the participant learns the pressed key is the right one; if the feedback is negative the participant knows the right key will be the opposite to the pressed one. This way, for a participant with 'perfect recall' the next time this image is displayed, i.e. its first repetition, the answer will be always correct. In this regard, a main element to be taken into account during task performance is the recalling of the learned association rule. Once the participant has learned the key associated to a specific image, the success likelihood at the next time the same image is presented will depend on the distance (in trials) to the previous presentation. Assuming that recall probability is described by exponential decay with the distance between two appearances of the same image, such repetition distance will be an important factor for task performance.

In order to assess the complexity of the task, let us consider the trial where all different scenes/situations (three scenes in the conditioning phase and six situations in the game phase) have repeated once (i.e. have appeared twice). For the sake of simplicity, and since the stimuli are sequences of images to be associated with the up-down arrow keys, we will refer to these scenes/situations as symbols. Figure 2S, main panel, shows the frequen-cy distribution of this 1st repetition trial for three and six different symbols. Dotted lines denote the interval containing the 90% of the probability of having all symbols repeated once: [6, 13] for the conditioning phase and [12, 33] for the game phase. On the other hand, the inset in Fig. 2S shows the frequency distribution of repetition distance for three and six different symbols. Dotted lines show the expected repetition distance: 3 for the conditioning phase and 6 for the game phase.

Since the game phase is more complicate than the conditioning training phase (six stimuli vs. three stimuli to be classified), its complexity will define the complexity of the proposed task. This way, for the game phase, in the 90% of random sequences composed of 33 symbols, all symbols will repeat once, so they would allow the learning the association rule under the assumption of 'perfect recall'. This conclusion is compatible with the expected stimulus length required to learn the rule: if the expected repetition distance is 6 and all six symbols must repeat at least once, the expected stimulus length for rule learning will be 36. Therefore, from this discussion, we can conclude that, in conditions of the Control group and being conservative, it would be expected to figure out the association rule during the game in around 30 trials.

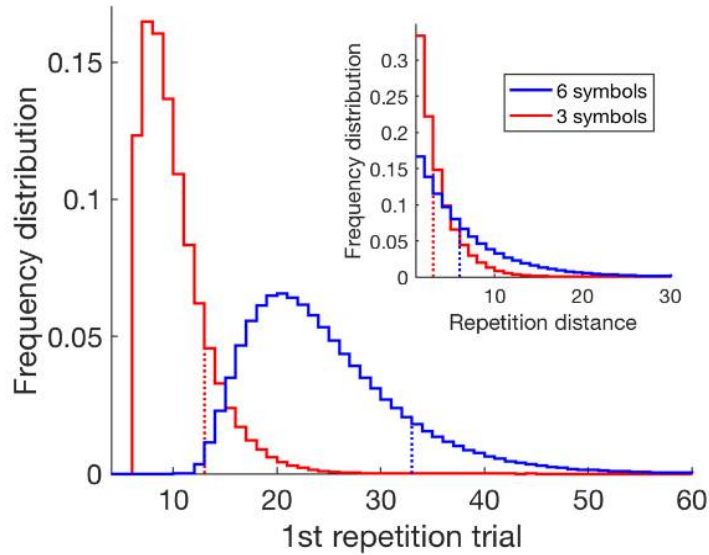

**Figure 2S. Evaluation of task complexity**. *Main panel*. Frequency distribution of the first trial where all symbols appeared repeated once. Dotted line denotes the interval of 1st repetition trials with probability equal to 0.9. *Inset.* Frequency distribution of repetition distance and expected repetition distance (dotted line). Red and blue lines stand for stimuli with three symbols (conditioning training phase) and six symbols (game phase).

**Stimulus difficulty**

In order to monitor the influence of intrinsic difficulty of stimuli over experiment performance we defined an index quantifying the difficulty to figure out the game association rule when a specific sequence of stimuli is displayed.

Let us consider a stimulus $s$ defined as a sequence of length $l$ made up of $m$ randomly distributed different symbols. We consider six possible symbols ($m \leq 6$), corresponding to the six dynamic situations displayed during the game. Let us denote by $d_{ij}$ the distance between the $j$-th repetition of the symbol $i$ ($i = 1, ..., m$) and its previous appearance, and by $r_i$ the number of repetitions of the symbol $i$ in the stimulus $s$ ($\sum_{i=1}^{m}(r_i + 1) = l$). We quantify the difficulty of the stimulus $s$ by evaluating the probability of learning the association rule when it is displayed in Control conditions. Assuming an exponential-decay recall process, the probability of recalling the correct association key for the specific symbol $i$ of the stimulus $s$ (and learnt after the first appearance of $i$ in $s$) will be

$$e^{-ad_{ij}} \qquad (1)$$

where $a$ is the decay constant. This way, the probability of answering correctly from the stimulus $s$ will be given by

$$p(s,a) = \prod_{k=1}^{l} p(k,s,a) \qquad (2)$$

where $p(k, s, a)$ stands for the probability of correct answer (right association key) at trial $k$, where the symbol $i$ ($i = 1, \dots, m$) is presented. Therefore, making explicit the probability for each one of the $m$ different symbols in $s$, $p(s, a)$ can be written as:

$$p(s, a) = \left(\frac{1}{2}\right)^m \prod_{i=1}^{m} \prod_{j=1}^{r_i} e^{-a d_{ij}} = \left(\frac{1}{2}\right)^m \prod_{i=1}^{m} e^{-a \sum_{j=1}^{r_i} d_{ij}} = \left(\frac{1}{2}\right)^m e^{-a \sum_{i=1}^{m} \sum_{j=1}^{r_i} d_{ij}} \qquad (3)$$

Applying logarithm to both sides of this expression we get

$$\log(p(s, a)) = \log\left(\left(\frac{1}{2}\right)^m e^{-a \sum_{i=1}^{m} \sum_{j=1}^{r_i} d_{ij}}\right) = -m \log 2 - a \sum_{i=1}^{m} \sum_{j=1}^{r_i} d_{ij} \qquad (4)$$

Note that the decay constant $a$ is a scale factor for the distance $d$, common to all stimuli. Therefore, without loss of generality, we will consider $a = 1$. On the other hand $m = 6$ for all stimuli, so the factor $m \log(2)$ will be a common onset. Therefore, from Eq. (s3) we define the *difficulty of stimulus s*, denoted by $D(s)$ as:

$$D(s) = \sum_{i=1}^{m} \sum_{j=1}^{r_i} d_{ij} \qquad (5)$$

This way the stimulus $s_1 = [1\,1\,1\,1\,1\,1\,1\,1]$ will have lower difficulty ($D(s) = 7.69$) than the stimulus $s_2 = [1\,2\,3\,4\,1\,2\,3\,4]$ ($D(s) = 18.77$). The stimulus difficulty defined in Eq. (5) strongly depends on the stimulus length $l$. Thus in order to define a difficulty index that can be compared among stimuli of different lengths: 1) it was generated a set of samples where each sample was made up of 100.000 randomly-generated stimuli of equal-length $l$, and each stimulus consist on a symbol sequence of length $l$ with six different symbols randomly distributed; lengths $l$ vary between 20 and 80 so 61 samples were generated. 2) For each sample, it was obtained the normalized frequency distribution of the difficulty, $f_l(c)$, where $l$ denotes the length of stimuli in the sample and $c$ is the class containing the difficulty. 3) Finally, given a stimulus $s$ of length $l$ and difficulty $D(s)$, it was quantized its *stimulus difficulty index DI(s)* by its cumulative frequency in $f_l(c)$, i.e:

$$DI(s) = \sum_{c=1}^{c_D} f_l(c) \qquad (6)$$

where $c_D$ denotes the class containing the difficulty $D(s)$ of the stimulus $s$.

The difficulty index allows us to show that there is no influence of the stimulus structure over experiment performance since no significant correlation exists between learning length and difficulty index (Fig. 3S).

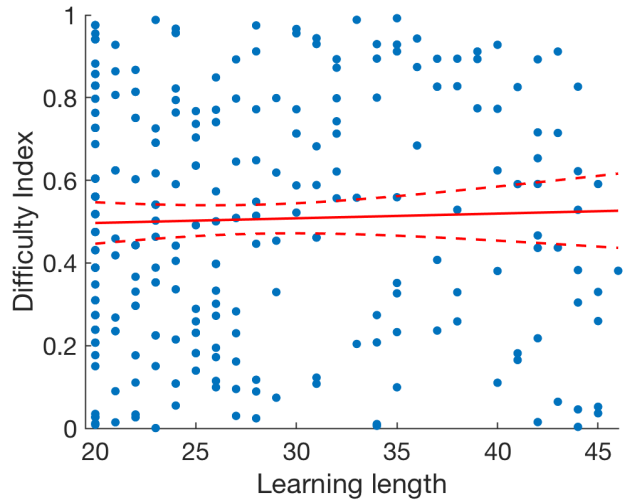

**Figure 3S. Stimulus difficulty vs. learning length.** No significant correlation is observed in the functional sample between stimulus difficulty index and learning length. *p-value* = 0.63.

**Individual stable performance**

It has been postulated that successfully triggering behavior in humans follows an exponential decay rate, related to the ability and motivation involved in a particular task [Fogg 2009]. Based on the stimulus difficulty evaluation, we can safely assume any participant in our experiment had enough ability to successfully carry it on. This means we can consider the ability as a fixed feature across the subjects, and reduce the outcome variability to the motivation dimension, thus dividing the sample between motivated individuals that correctly performed the task, showing a stable performance, and unmotivated individuals who did not. A potential indicator of this latter group is the appearance of a sample fraction (18%) that does not fit for proportional hazards between experiment and gender in the Cox model, which in addition corresponds to participants with learning lengths greater than 46 (Fig. 4S).

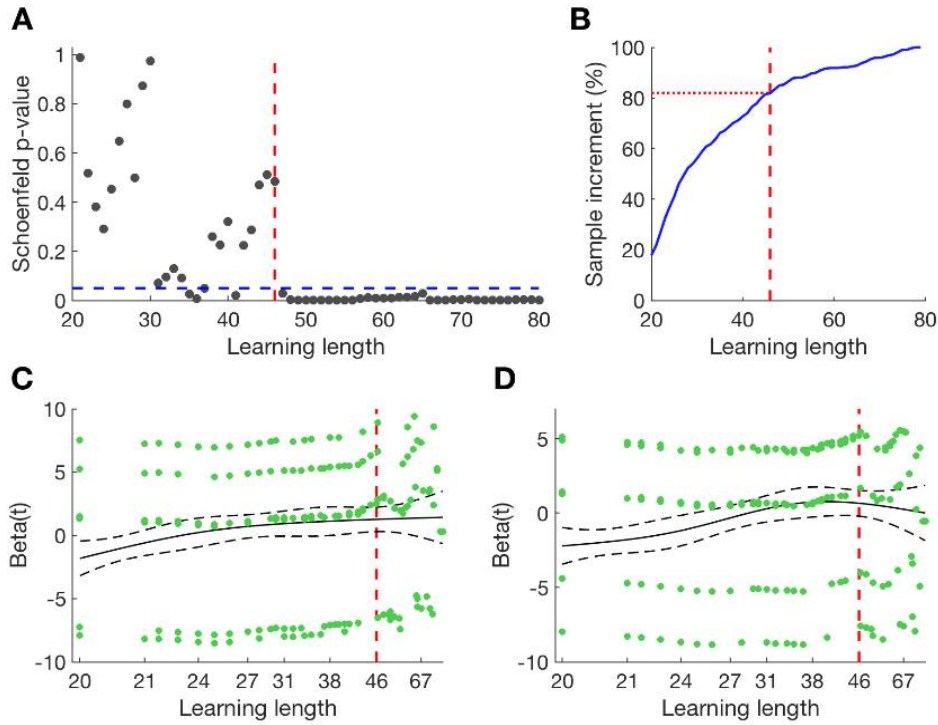

**Figure 4S. Proportional hazard ratio model assumptions.** Schoenfeld test always rejects Cox assumptions satisfaction hypothesis (proportional hazard assumption) when considering individuals with learning length > 46 (blue dashed line at 0.05 and red dashed line at 46). B) Smooth increment in sample size from learning length equal to 46 shows the sudden drop in the Schoenfeld p-value is not caused by sampling in the Schoenfeld test (82% of the sample corresponds to learning lengths ≤ 46). C) and D). Scaled Schoenfeld residuals against transformed time. Beta(t) stands for regression coefficients of interactions terms between gender and experiment factors. The diagnostic plots show a fitted spline of 4 degrees of freedom (solid line) and a 95% confidence interval (dashed lines). The parallel structure expected from the proportional hazard assumption is lost for learning lengths greater than 46, indicating violation of non-proportionality from this value.

In view of these results, we explored more in-depth the performance stability of each participant to assess the existence of a 'demotivated' group pointed by the collapse of the proportional hazard assumption. In a trial-error procedure, when a certain stimulus appears for the first time the subject must randomly respond, so success probability to the first appearance is 0.5. In this context, if the same stimulus is repeated immediately after this first appearance (i.e. at the next trial) the subject will always guess the correct answer. Thus, success probability for the first repetition would be 1. However, if this first repetition occurs at a certain distance from the first occurrence (in number of trials), its success probability could be affected by various factors, such as distraction, forgetting, etc. The same can be applied to the following repetitions of the stimulus. We only use the first two repetitions, since, according to the simplicity of the task (Fig 2S), a participant who is properly performing the task should have achieved a high learning rate at this moment. Taking the six stimuli together, we calculated the success frequency $f_i(d)$ of the entire sample at each repetition distance $d$, where $i$ =1, 2 stand for the first and second repetition respectively. For each subject its *response* probability to the stimulus $j$ which appears at a distance $t_j$ from its previous appearance, $rp_{i,j}$, will be $f_i(t_j)$ if he responded correctly and 1-$f_i(t_j)$ if he failed. This way, the *repetition* probability $rp_i$ for each individual will be:

$$rp_i = \prod_{j=1}^{6} rp_{i,j} \qquad (7)$$

assuming the independence of each response to a certain stimulus. If a subject is performing the task properly, high probabilities of at last one repetition will be expected. On the contrary a subject not responding properly will have 1st and 2nd repetition probabilities around or lower than 0.5[6], the expected value for random answer probabilities. Density plot of 1st vs. 2nd repetition probabilities shown in Fig 5S corroborates previous Cox regression analysis, displaying highest density of subjects with learning lengths ≤ 46 in good-performance areas whereas subjects with learning lengths >46 corresponds to people who are not correctly performing the task. The detailed distribution of these two populations according to experimental group and gender shows that most subjects with learning lengths ≤ 46 showed at least either the 1st or the 2nd repetition probabilities above randomness (Fig 6S) while subjects with learning lengths > 46 lie around randomness (Fig 7S).

Putting together Cox regression analysis and repetition probabilities we conclude that there exist two populations characterized by their performance during the game phase, showing that those participants with learning length below or equal to 46 exhibit a stable performance, which ensures an adequate and valid realization of the experiment.

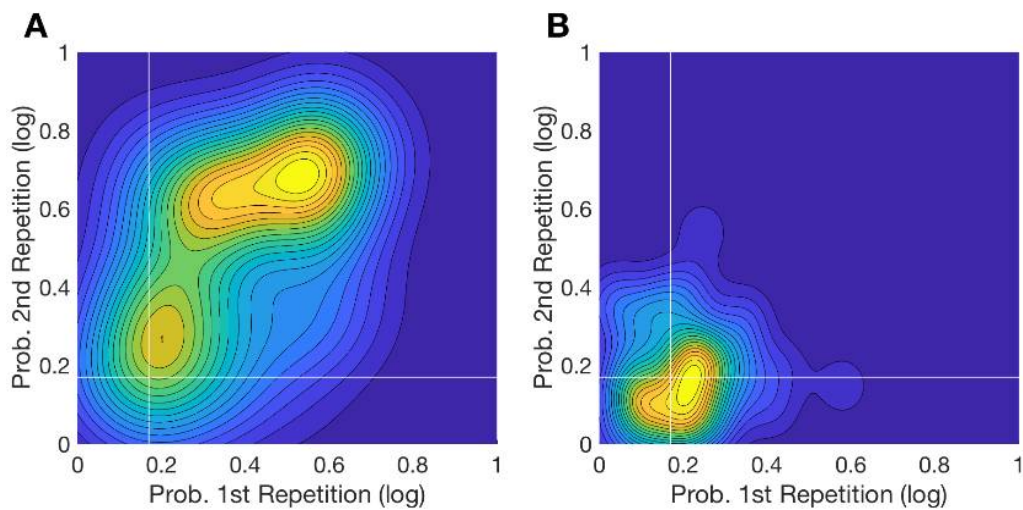

**Figure 5S. Density plot of repetition probability.** A) Density plot of 1st vs. 2nd repetition probabilities for subjects with learning lengths ≤ 46 (A) and for subjects with learning lengths > 46 (B). Vertical and horizontal scales are logarithmic (log(1 + sqrt(probability)) is represented). White vertical and horizontal lines represent repetition probabilities for random answer.

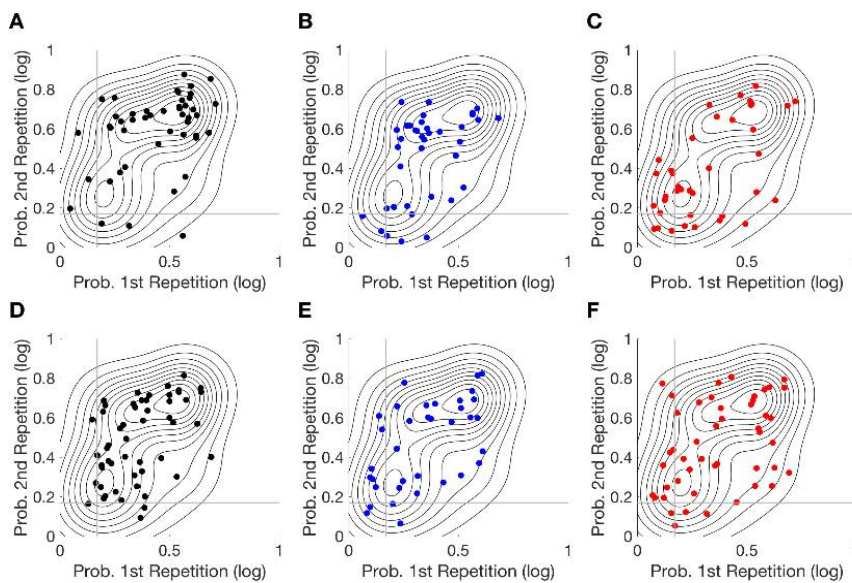

**Figure 6S. Distribution of repetition probability according to experiment and gender for Learning Length <= 46.** A, B, C) Distribution for Favored, Control, and Hampered men respectively. D, E, F) Distribution for Favored, Control, and Hampered women respectively.

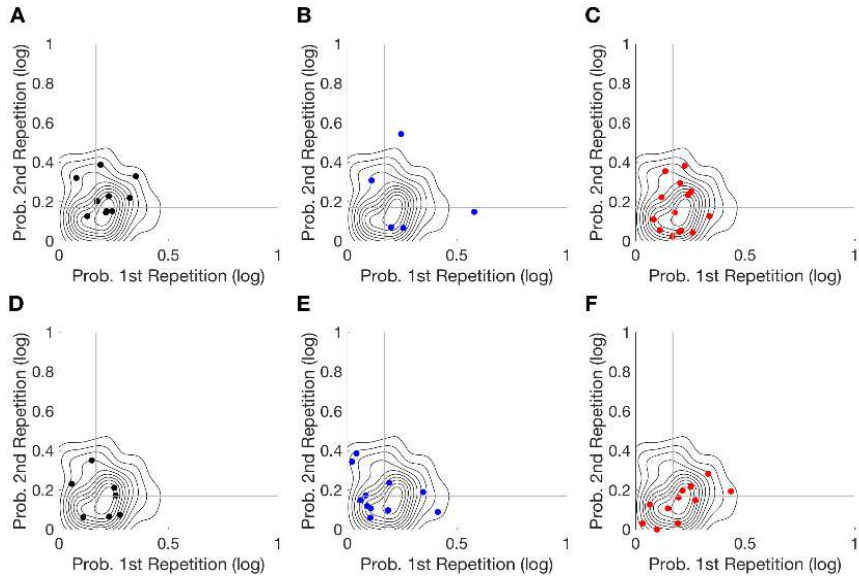

**Figure 7S. Distribution of repetition probability according to group and gender for Learning Length > 46.** A, B, C) Distribution for Favored, Control, and Hampered men respectively. D, E, F) Distribution for Favored, Control, and Hampered women respectively.

## Mathematical model

### Analytical model of the population learning process

Along this text it will be assumed that stimulus, recall process, and probabilities refer to the game phase for the Favored, Control, and Hampered groups.

Let us consider the random variable $S$ denoting the success in the participant's answer when a dynamic situation is displayed. Therefore, $S = 1 / S = 0$ stands for correct/wrong answer respectively, i.e. the pressed key matches / does not match with the displayed dynamic situation according to the underlying game association rule.

On the other hand we consider that a recall process exists, described by a random variable $R$, with $R = 1$ or $0$ if the subject recalls or not the corresponding stimulus. Assuming the recall process is present during the experiments, the probability of getting a correct answer when the stimulus $i$ appeared $t$ time units ago will be

$$P(S = 1 | i \text{ appeared } t \text{ time ago}) \equiv P_{i,t}(S = 1) = P_{i,t}(S = 1 \cap R) =$$
$$= P(S = 1 | R = 1)P_{i,t}(R = 1) + P(S = 1 | R = 0)P_{i,t}(R = 0) \quad (8)$$

The recall probability will be

$$P_{i,t}(R = 1) = e^{-at} \quad (9)$$

where $a$ is the recalling decay rate, whereas the probability of non-recalling will be

$$P_{i,t}(R = 0) = 1 - P_{i,t}(R = 1) = 1 - e^{-at} \quad (10)$$

On the other hand we assume that when the subject recalls, the answer will be always correct, so

$$P(S = 1 | R = 1) = 1 \quad (11)$$

and when the subject does not recall, the answer will be random. Thus

$$P(S = 1 | R = 0) = \frac{1}{2} \quad (12)$$

This way, from Eq. (8), the probability of getting the right answer by recalling the key associated to the stimulus $i$ (learnt $t$ time ago) will be

$$P_{i,t}(S = 1) = e^{-at} + \frac{1}{2}(1 - e^{-at}) = \frac{1}{2} + \left(1 - \frac{1}{2}\right)e^{-at} = \frac{1 + e^{-at}}{2} \quad (13)$$

Let us assume now that 1) time is discrete, measured in terms of *trials* and 2) the stimulus *i* has appeared at the trial *T*. Thus the probability of right answer will depend on when the stimulus *i* previously appeared, since the association rule will be learnt once it appears, but its recalling will depend on the distance (in trials) between its previous and present appearances.

*The model*

Let us assume the involved recalling process is only considered when a single stimulus appears up to three times. Thus if this stimulus appears four o more times it is assumed the corresponding association will be always recalled and the probability of successful answer will be 1.

In order to sytematize the model, let us define the random variable $X_i$ as the number of times the stimulus *i* appeared in the interval [1, …, T-1]. Thus we can write the overall probability of success when stimulus *i* has appeared at the trial *T* as:

$$P_i(S = 1) = P_i(S = 1 \cap X_i = 0) + P_i(S = 1 \cap X_i = 1) + P_i(S = 1 \cap X_i = 2) + P_i(S = 1 \cap X_i \geq 3) \quad (14)$$

where subindex *i* denotes probabilities referred to the stimulus *i*.

Let us analyze separately the four terms in the right side of Eq. (14).

If the stimulus hasn't appeared yet in [1, …, T-1] (so it appears for the first time at trial *T*), there's no prior information, so the subject will press one of the two keys randomly, i.e. $P(S = 1 | X_i = 0) = 1/2$, and

$$P_i(S = 1 \cap X_i = 0) = P_i(S = 1 | X_i = 0) \cdot P_i(X_i = 0) = \frac{1}{2}\left(\frac{5}{6}\right)^{T-1} \quad (15)$$

If the stimulus has appeared once previously, then

$$P_i(S = 1 \cap X_i = 1) = \sum_{t=1}^{T-1} P_i(S = 1 | X_i = 1) \cdot P_i(X_i = 1)$$

On the one hand, according to the binomial distribution

$$P_i(X_i = 1) = \left(\frac{1}{6}\right)\left(\frac{5}{6}\right)^{T-2}$$

On the other hand, if the stimulus *i* has appeared at the trial *t* then

$$P_i(S = 1 | X_i = 1) = P_i(S = 1 | i \text{ appeared } T - t \text{ time ago}) \equiv P_{i,T-t}(S = 1)$$

so

$$P_i(S = 1 \cap X_i = 1) = \sum_{t=1}^{T-1} P_{i,T-t}(S = 1 | X_i = 1) \cdot P_i(X_i = 1) = \left(\frac{1}{6}\right)\left(\frac{5}{6}\right)^{T-2} \sum_{t=1}^{T-1} \frac{1 + e^{-a(T-t)}}{2} =$$

$$= \left(\frac{1}{6}\right)\left(\frac{5}{6}\right)^{T-2} \left(\frac{T-1}{2} + \frac{1}{2}\sum_{t=1}^{T-1} e^{-a(T-t)}\right) \quad (16)$$

where

$$\sum_{t=1}^{T-1} e^{-a(T-t)} = \frac{1 - e^{-a(T-1)}}{e^a - 1}$$

Substituting back in Eq. (16)

$$P_i(S = 1 \cap X_i = 1) = \left(\frac{1}{6}\right)\left(\frac{5}{6}\right)^{T-2} \frac{1}{2}\left[T - 1 + \left(\frac{1 - e^{-a(T-1)}}{e^a - 1}\right)\right] \quad (17)$$

Let us consider now that the stimulus *i* appears twice in the interval $t \in [1, …, T - 1]$, at trials $t_1$ and $t_2$. Then

$$P_i(S = 1 \cap X_i = 2) = \sum_{t_1=1}^{T-1} \sum_{t_2=t_1+1}^{T-1} P_i(S = 1 \mid X_i = 2) \cdot P_i(X_i = 2) \quad (18)$$

The probability $P_i(X_i = 2)$ of the stimulus $i$ appears twice in the interval $t \in [1, \ldots, T-1]$ is given by

$$P_i(X_i = 2) = \frac{(T-1)(T-2)}{2} \left(\frac{1}{6}\right)^2 \left(\frac{5}{6}\right)^{T-3} \quad (19)$$

On the other hand the probability of recalling the stimulus association depends only on the distance between $T$ and $t_2$, the last trial where the stimulus appeared. Therefore

$$P_i(S = 1 \mid X_i = 2) = P_i(S = 1 \mid i \text{ appeared } T - t_2 \text{ time ago}) \equiv P_{i,T-t_2}(S = 1)$$

so

$$P_i(S = 1 \cap X_i = 2) = \sum_{t_1=1}^{T-1} \sum_{t_2=t_1+1}^{T-1} P_i(S = 1 \mid X_i = 2) \cdot P_i(X_i = 2)$$

$$= \left(\frac{1}{6}\right)^2 \left(\frac{5}{6}\right)^{T-3} \sum_{t_1=1}^{T-1} \sum_{t_2=t_1+1}^{T-1} \left(\frac{1 + e^{-b(T-t_2)}}{2}\right) \quad (20)$$

Note that the recalling constant is now denoted by $b$. We develop the sum to obtain:

$$\sum_{t_1=1}^{T-1} \sum_{t_2=t_1+1}^{T-1} \left(\frac{1 + e^{-b(T-t_2)}}{2}\right) = \frac{1}{2} \sum_{t_1=1}^{T-1} \sum_{t_2=t_1+1}^{T-1} 1 + \frac{e^{-bT}}{2} \sum_{t_1=1}^{T-1} \sum_{t_2=t_1+1}^{T-1} e^{bt_2} =$$

$$= \frac{(T-1)(T-2)}{4} + \frac{1}{2(e^b - 1)} \left[(T-1) - \left(\frac{1 - e^{-b(T-1)}}{1 - e^{-b}}\right)\right] \quad (21)$$

Substituting the Eq. (21) into Eq. (20) we finally get

$$P_i(S = 1 \cap X_s = 2) =$$

$$= \left(\frac{1}{6}\right)^2 \left(\frac{5}{6}\right)^{T-3} \left[\frac{(T-1)(T-2)}{4} + \frac{1}{2(e^b - 1)} \left[(T-1) - \left(\frac{1 - e^{-b(T-1)}}{1 - e^{-b}}\right)\right]\right] \quad (22)$$

The remaining term in Eq. (14) refers to the situation where the stimulus has appeared three or more times

$$P_i(S = 1 \cap X_i \geq 3) = P_i(S = 1 \mid X_i \geq 3) \cdot P_i(X_i \geq 3) \quad (23)$$

As the model assumes that the association rule will be always recalled after the stimulus $i$ appears three or more times in the interval $t \in [1, \ldots, T-1]$, then $P_i(S = 1 \mid X_i \geq 3) = 1$ so

$$P_i(S = 1 \cap X_i \geq 3) = P_i(X_i \geq 3) = 1 - P_i(X_i = 0) - P_i(X_i = 1) - P_i(X_i = 2) =$$

$$= 1 - \left(\frac{5}{6}\right)^{T-1} - \left[(T-1)\left(\frac{1}{6}\right)\left(\frac{5}{6}\right)^{T-2}\right] - \left[\frac{(T-1)(T-2)}{2} \left(\frac{1}{6}\right)^2 \left(\frac{5}{6}\right)^{T-3}\right] \quad (24)$$

Finally, going back to Eq. (14):

$$P_i(S = 1) = P_i(S = 1 \cap X_i = 0) + P_i(S = 1 \cap X_i = 1) + P_i(S = 1 \cap X_i = 2) + P_i(S = 1 \cap X_i \geq 3)$$

and substituting the Eqs. (15), (17), (22) and (24) on it, we obtain the probability of successful answer when the stimulus $i$ appears at trial $T$:

$$P_i(S = 1) = 1 - \frac{1}{2}\left(\frac{5}{6}\right)^{T-1} - \frac{1}{2}\left[(T-1) - \left(\frac{1 - e^{-a(T-1)}}{e^a - 1}\right)\right]\left(\frac{1}{6}\right)\left(\frac{5}{6}\right)^{T-2}$$

$$-\frac{1}{4}\left[(T-1)(T-2)-\frac{2}{(e^b-1)}\left[(T-1)-\left(\frac{1-e^{-b(T-1)}}{1-e^{-b}}\right)\right]\right]\left(\frac{1}{6}\right)^2\left(\frac{5}{6}\right)^{T-3} \quad (25)$$

*Control group*

The experiment conducted for the Control group consists on a conditioning phase 'detached' from the game phase, since no conditioning SM is displayed. This way the probability of successful answer during the game will be the same for the six stimuli presented. Then we can write the overall probability of successful answer at the game trial $T$ as

$$P(S=1) = \sum_{i=1}^{6} P_i(S=1) \cdot P(i \text{ appears at trial } T) \quad (26)$$

Since at any single trial all stimuli are equiprobable then $P(i \text{ appears at trial } T) = 1/6$. Besides $P_i(S=1)$ does not depend on the stimulus, so

$$P(S=1) = P_i(S=1) \sum_{i=1}^{6} \frac{1}{6} = P_i(S=1)\frac{1}{6}6 = P_i(S=1)$$

and, from Eq. (25), the probability of successful answer at trial $T$ for the Control group will be:

$$P_C(S=1) = 1 - \frac{1}{2}\left(\frac{5}{6}\right)^{T-1} - \frac{1}{2}\left[(T-1)-\left(\frac{1-e^{-a(T-1)}}{e^a-1}\right)\right]\left(\frac{1}{6}\right)\left(\frac{5}{6}\right)^{T-2}$$

$$-\frac{1}{4}\left[(T-1)(T-2)-\frac{2}{(e^b-1)}\left[(T-1)-\left(\frac{1-e^{-b(T-1)}}{1-e^{-b}}\right)\right]\right]\left(\frac{1}{6}\right)^2\left(\frac{5}{6}\right)^{T-3} \quad (27)$$

*Favored group*

According to the hypothesis, for Favored group the DMs of the game, here denoted by $i$ = 1 and 2, were previously learned during the conditioning, after the SM was learnt. Therefore, when DMs appear during the game, the probability of successful answer will be always equal to 1.

This way the successful answer probability during the game will be 1 for stimuli $i$ = 1 and 2, and will be given by Eq. (25) for the remaining stimuli $i$ = 3, ..., 6. Thus considering again that, at any single trial, all stimuli are equiprobable, we can write the overall probability of successful answer at the game trial $T$ for the Favored group as

$$P_F(S=1) = \sum_{i=1}^{6} P_i(S=1) \cdot P(i \text{ at } T) = \frac{1}{6}\sum_{i=1}^{6} P_i(S=1)$$

$$= \frac{1}{6}\left(1+1+\sum_{i=3}^{6} P_i(S=1)\right) = \frac{1}{6}\left(2+4P_i(S=1)\right) \quad (28)$$

Substituting Eq. (25) we obtain

$$P_F(S=1) = 1 - \frac{1}{3}\left(\frac{5}{6}\right)^{T-1} - \frac{1}{3}\left[(T-1)-\left(\frac{1-e^{-a(T-1)}}{e^a-1}\right)\right]\left(\frac{1}{6}\right)\left(\frac{5}{6}\right)^{T-2}$$

$$-\frac{1}{6}\left[(T-1)(T-2)-\frac{2}{(e^b-1)}\left[(T-1)-\left(\frac{1-e^{-b(T-1)}}{1-e^{-b}}\right)\right]\right]\left(\frac{1}{6}\right)^2\left(\frac{5}{6}\right)^{T-3} \quad (29)$$

*Hampered group*

Following the hypothesis, for Hampered group the DMs of the game phase, again denoted by $i$ = 1 and 2, were previously wrongly learned during the conditioning phase, i.e. the corresponding arrow key for the DMs is the opposite to that learned for the conditioning SM. Therefore the probability of successful answer when they appear for the first time during the game will be equal to 0. When they appear for the second and successive times the new association rule will be learnt and the probability of successful answer will be given by Eq. (18). More in detail, let us consider the Eq. (14) for $i$ = 1 and 2:

$$P_{1,2}(S = 1) = P_{1,2}(S = 1 \cap X_i = 0) + P_{1,2}(S = 1 \cap X_i = 1) + P_{1,2}(S = 1 \cap X_i = 2) + P_{1,2}(S = 1 \cap X_i \geq 3)$$

On the one hand

$$P_{1,2}(S = 1 \cap X_i = 0) = P_{1,2}(S = 1 | X_i = 0) \cdot P_{1,2}(X_i = 0) = 0 \cdot \left(\frac{5}{6}\right)^{T-1} = 0 \quad (30)$$

On the other hand the remaining terms $P_{1,2}(S = 1 \cap X_i = 1)$, $P_{1,2}(S = 1 \cap X_i = 2)$, and $P_{1,2}(S = 1 \cap X_i \geq 3)$ will be given respectively by Eq. (17), (22) and (24), so we get

$$P_{1,2}(S = 1) = 1 - \left(\frac{5}{6}\right)^{T-1} - \frac{1}{2}\left[(T-1) - \left(\frac{1 - e^{-a(T-1)}}{e^a - 1}\right)\right]\left(\frac{1}{6}\right)\left(\frac{5}{6}\right)^{T-2}$$

$$- \frac{1}{4}\left[(T-1)(T-2) - \frac{2}{(e^b - 1)}\left[(T-1) - \left(\frac{1 - e^{-b(T-1)}}{1 - e^{-b}}\right)\right]\right]\left(\frac{1}{6}\right)^2\left(\frac{5}{6}\right)^{T-3} \quad (31)$$

Considering then that $P_i(S = 1)$ when $i$ = 3, ..., 6 is described by Eq. (25), we can write the overall probability of successful answer at the game trial $T$ for the Hampered group as

$$P_H(S = 1) = \sum_{i=1}^{6} P_i(S = 1) \cdot P(i \text{ at } T) = \frac{1}{6}\sum_{i=1}^{6} P_i(S = 1) = \frac{1}{6}\left(\sum_{i=1}^{2} P_i(S = 1) + \sum_{i=3}^{6} P_i(S = 1)\right)$$

so substituting and simplifying we obtain that

$$P_H(S = 1) = 1 - \frac{2}{3}\left(\frac{5}{6}\right)^{T-1} - \frac{1}{2}\left[(T-1) - \left(\frac{1 - e^{-a(T-1)}}{e^a - 1}\right)\right]\left(\frac{1}{6}\right)\left(\frac{5}{6}\right)^{T-2}$$

$$- \frac{1}{4}\left[(T-1)(T-2) - \frac{2}{(e^b - 1)}\left[(T-1) - \left(\frac{1 - e^{-b(T-1)}}{1 - e^{-b}}\right)\right]\right]\left(\frac{1}{6}\right)^2\left(\frac{5}{6}\right)^{T-3} \quad (32)$$

In order to summarize the previous equations, we can write the successful answer probabilities at trial $T$ for Favored, Control, and Hampered groups as:

$$P_F(S = 1) = 1 - \frac{1}{3}\left(\frac{5}{6}\right)^{T-1}[1 + \alpha(a,T) + \beta(b,T)] \quad (33)$$

$$P_C(S = 1) = 1 - \frac{1}{2}\left(\frac{5}{6}\right)^{T-1}[1 + \alpha(a,T) + \beta(b,T)] \quad (34)$$

$$P_H(S = 1) = 1 - \frac{1}{2}\left(\frac{5}{6}\right)^{T-1}\left[\frac{4}{3} + \alpha(a,T) + \beta(b,T)\right] \quad (35)$$

where

$$\alpha(a,T) = \left[(T-1) - \left(\frac{1 - e^{-a(T-1)}}{e^a - 1}\right)\right]\left(\frac{1}{5}\right) \quad (36)$$

$$\beta(b,T) = \frac{1}{2}\left[(T-1)(T-2) - \frac{2}{(e^b - 1)}\left[(T-1) - \left(\frac{1 - e^{-b(T-1)}}{1 - e^{-b}}\right)\right]\right]\left(\frac{1}{5}\right)^2 \quad (33)$$

**Computational model of the individual learning performance**

To assess individual learning performance on the basis of the time compaction hypothesis, a set of $10^5$ virtual participants were considered for each experimental group. For each virtual subject, an 80-trials length sequence of random 'stimuli' was generated (a number between 1 and 6 corresponding to one of the six displayed dynamic situations). According to the model's assumption, the response to the first appearance of a given stimulus de-pends on the previous conditioning: for Control group the answer will always be random, whereas Favored and Hampered group will always guess right or fail the first appearance of those stimuli corresponding to DMs (dynamic matching situations). For the subsequent trials, the virtual participant will answer according to a exponential recalling process until the learning has been attained, which is assumed at fourth or later appearance of the stimulus. Therefore the model assumes recalling can only happen in the second and third occurrences of the stimulus, so for these simulation trials, the probabilities of recalling will be respectively:

$$p(\text{recall at } T_2) = e^{-a(T_2-T_1)} \quad (34)$$

$$p(\text{recall at } T_3) = e^{-b(T_3-T_2)} \quad (35)$$

where *a* and *b* are the recalling decay rates, and $T_i$ stands for the trials in which the *i*-th appearance of such a stimulus occurs. Following a Monte-Carlo simulation, at each trial $T_{\{2,3\}}$ a random uniform number between 0 and 1 is generated: if it is lower than the recalling probabilities (34) or (35), the virtual subject's answer will be correct; if not, the virtual response will be random. For four or more occurrences of a certain stimulus it is assumed the answer is always correct. This process is repeated for every participant until the learning criteria is fulfilled (18 successes in the last 20 trials).